\documentclass[twocolumn,prl]{revtex4}

\usepackage{tikz,pgf}
\usepackage{amsmath,amsfonts,amssymb,array}
\usepackage{color}
\usepackage{graphicx}

\def \be {\begin{enumerate}}
\def \ee {\end{enumerate}}
\def \beq {\begin{equation}}
\def \eeq {\end{equation}}
\def \ba {\begin{eqnarray}}
\def \ea {\end{eqnarray}}
\def \ban {\begin{eqnarray*}}
\def \ean {\end{eqnarray*}}
\def \bfl {\begin{flalign*}}
\def \efl {\end{flalign}}
\def \bsp {\begin{split}}

\def \l {\left}
\def \r {\right}

\newcommand{\ketbrad}[1]{|#1\rangle\!\langle #1|}

\newcommand{\mc}[1]{{\mathcal #1}}
\newcommand{\mbb}[1]{{\mathbb #1}}
\newcommand{\mean}[1]{\langle#1\rangle}

\newcommand{\Mean}[1]{\left\langle#1\right\rangle}

\newcommand{\tr}[1] {\mbox{tr} \left\{ #1 \right \}}

\begin{document}
\title{A noise inequality for classical forces}
\author{D. Kafri}
\affiliation{Joint Quantum Institute/NIST, College Park, MD, USA}
\author{J. M. Taylor}
\affiliation{Joint Quantum Institute/NIST, College Park, MD, USA}
\begin{abstract}
Lorentz invariance requires local interactions, with force laws such as the Coulomb interaction arising via virtual exchange of force carriers such as photons. Many have considered the possibility that, at long distances or large mass scales, this process changes in some way to lead to classical behavior.  Here we hypothesize that classical behavior could be due to an inability of some force carriers to convey entanglement, a characteristic measure of nonlocal, quantum behavior.  We then prove that there exists a local test that allows one to verify entanglement generation, falsifying our hypothesis.  Crucially, we show that noise measurements can directly verify entanglement generation.  This provides a step forward for a wide variety of experimental systems where traditional entanglement tests are challenging, including entanglement generation by gravity alone between macroscopic torsional oscillators. 
\end{abstract}

\maketitle
All known interactions are, at a fundamental level, local.  That is, any Lorentz-invariant Lagrangian density must have only local terms, such as particle-particle and particle-gauge field interactions.  From these local interactions, non-local, long-range force `laws' such as the Coulomb interaction can emerge in a natural way via the `virtual' fluctuations of gauge fields \cite{Weinberg1995,Srednicki2007,Zee2010}.  However, while this method is extremely successful in understanding small-scale behavior, it leaves large questions about falsifiability -- how do you test for virtual particles?  

Here we propose an inequality associated with force laws, whose violation indicates that the associated force law is necessarily transmitting quantum information.  Our approach is guided by the `gold standard' for quantum behavior: Bell's inequality\cite{Bell1964, Aspect1981, Clauser1969}.  Consequently, we must start by revisiting the meaning of classical behavior. One notion precludes large spatial superpositions at macroscopic scales, which may be explained by decoherence due to a general quantum environment \cite{Caldeira1983,Joos1985} or through interactions with a noisy gravitational field \cite{Diosi1987,Diosi1989} (for a specific experimental proposal testing such ideas, see \cite{vanWezel2012}.)  
 We choose instead to work in the spirit of~\cite{Gell-Mann1993} and define a classical interaction as one that \emph{cannot entangle} and yet leads to the expected classical equations of motion through, e.g., Ehrenfest's theorem\cite{Ehrenfest1927}.  In many respects, our work is similar to previous efforts to work with semiclassical gravity and their quantum consequences, such as~\cite{Yang2013}. Under this definition, classical behavior involving long range forces emerges naturally from interactions between the intermediary gauge field and a large quantum environment. We provide a simple model for how this comes about, and derive a locally observable consequence of this definition of classicality.  Finally, we conclude with a proposed experiment for testing the ability to entangle massive objects via gravity.  Most curiously, our approach suggests that, as a substitution for Bell's inequality, one can verify entanglement generation with a two step approach: first, measure the linear response as a ``witness'' of the coupling strength, then measure the noise and compare to the information transfer suggested by the witness.  This enables a noise-based test of entanglement generation which may be easier to confirm in laboratory settings.  


By using a force carrier (FC) intermediary, we first derive an effective nonlocal interaction corresponding to Hamiltonian $H_\Sigma = H_a + H_b + H_{ab}$. Setting $\hbar = 1$ from this point onward, $H_a$ and $H_b$ are local Hamiltonian terms acting on (separate) quantum systems, $a$ and $b$, and $H_{ab} = A B$ is a product of operators acting on each. The FC is an ancillary harmonic oscillator with canonical coordinates $x$ and $p$, and interacts individually with each system through the chain of unitary evolutions \cite{Milburn2000}
\beq
e^{i \sqrt{\tau} p B} e^{i \sqrt{\tau} x A} e^{-i \sqrt{\tau} p B} e^{-i \sqrt{\tau} x A} = e^{-i\tau A B}\,.
\eeq
Heuristically picturing the FC as traveling between $a$ and $b$, in this way evolution under $H_{ab}$ can be implemented through local interactions. Following this by the local unitary $e^{-i \tau(H_a + H_b)}$ completes the evolution for small time step $\tau$. Repeating this process $n$ times while keeping $t = n \tau$ fixed, by the Trotter formula we obtain the overall desired evolution $\exp(-i t H_{\Sigma}) = \lim_{n \rightarrow \infty} \left(  e^{-i A B t/n} e^{-i (H_a + H_b)t/n} \right)^{n}$. Notice that since the force carrier is completely uncorrelated with the systems at the end of the time step, we may then trace it out. This effective interaction is the underlying mechanism describing geometric phase gates \cite{Molmer1999,Milburn2000, Garcia-Ripoll2005}.

Within our model we may impose our classical interaction constraint -- no entanglement can be generated by the interaction -- by introducing as a conceptual aid a `screen' operation on the FC. The screen acts like a weak measurement in the middle of the infinitesimal time step, whose strength will determine whether the FC is able to entangle $a$ and $b$. Letting $\mc{S}$ be the trace-preserving, completely positive map \cite{Nielsen2000} representing the screen's action on the force carrier, the complete evolution (Fig.~1) is described by the superoperator
\begin{figure}
\includegraphics[width=0.45\textwidth]{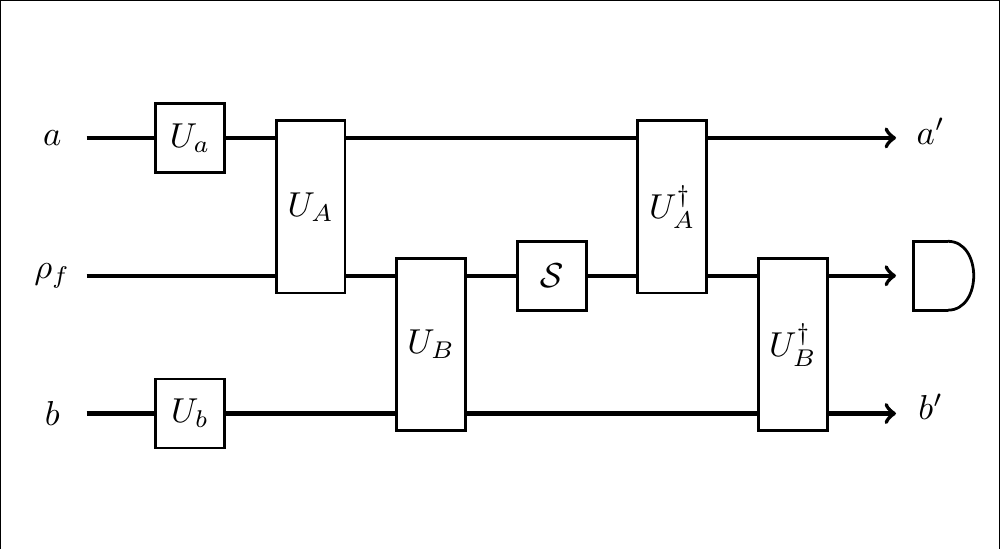}
\caption{Quantum circuit schematic of infinitesimal time step implementing a virtual particle exchange between two systems, with a potential entanglement-reducing screen $\mc{S}$, where $U_a = \exp(-i t H_a)$, $U_b = \exp(-i t H_b)$, $U_A = \exp(-i \sqrt{ t} A x )$, $U_B = \exp(-i \sqrt{ t} B p )$}. 
\end{figure} 
\beq
\label{circuit}
\mc{V}_t = \mc{C}_{-B p \sqrt{t}}\,  \mc{C}_{-A x \sqrt{t}}\, \mc{S} \, \mc{C}_{B p \sqrt{t}}\, \mc{C}_{A x \sqrt{t}} \, \mc{C}_{(H_a+H_b)t}\,,
\eeq
where $\mc{C}_{X}(\rho) = e^{-i X} \rho e^{i X}$ are superoperators associated with unitary evolution and the products above denote composition. As expected, when $\mc{S}$ is the identity, the Trotter limit is just $\mc{C}_{H_\Sigma t}$. 

The superoperator of Equation \eqref{circuit} acts non-trivially on the FC, and hence cannot determine the dynamics of $a$ and $b$ alone. To do this we take the Markovian limit, i.e. we suppose that the FC starts in the same density matrix, $\rho_f$, between each time step. This is equivalent to saying that the FC has no ``memory'' of its interactions with $a$ and $b$, and that it is drawn out of a large reservoir of identical systems. We can then trace out the FC, yielding the reduced infinitesimal propagator,
\beq
\label{circuit2}
\mc{V}^{red}_{t}(\rho_{a b}) = \mbox{tr}_{f} \l\{ \mc{V}_{t}(\rho_{a b}\otimes \rho_f)\r\}\,.
\eeq
Finally, we take the take the Trotter limit to produce the true time evolution superoperator,
\beq
\label{trotter}
 e^{\mc{L} t} = \lim_{n\rightarrow \infty} \l(\mc{V}^{red}_{t/n} \r)^n\,,
\eeq
with corresponding generator $\mc{L} \equiv \partial_t (\mc{V}^{red}_t) |_{t = 0}$\cite{Lindblad1976}. 

Before showing how our model produces the expected dynamics, we first derive some properties of the screen $\mc{S}$ based on physical considerations. Our first requirement is that the limit \eqref{trotter} converges, i.e., that the terms of order $\sqrt{t}$ in the Taylor expansion of $\mc{V}^{red}_t$ vanish. To see when this is the case, we use the Baker-Campbell-Hausdorff formula
\beq
\label{BCH}
e^{i X} \rho e^{-i X} = \rho + \frac{1}{1!}[i X,\rho] + \frac{1}{2!} [i X, [i X, \rho]]+...
\eeq
Substituting into Equations \eqref{circuit} and \eqref{circuit2}, we see (in the appendix) that the $\sqrt{t}$ terms vanish if and only if $\mc{S}$ preserves the quadratures of the FC,
\beq
\label{efest1}
\Mean{ \mc{S}^\dagger(x)-x}_f = \Mean{ \mc{S}^\dagger(p)-p}_f = 0\,,
\eeq
where $\Mean{O}_f = \tr{\rho_f\,O}$ and $\mc{S}^\dagger$ is the Hilbert-Schmidt Hermitian adjoint of $\mc{S}$, corresponding to the Heisenberg picture, $ \tr{\rho_f S^\dagger(O)} \equiv \tr{S(\rho_f)O}$.

Our second physical requirement is that classical mechanics still emerge from the model.  In this setting, this requires that all canonical variables obey Ehrenfest's theorem\cite{Ehrenfest1927}. If $A'$ and $B'$ represent such variables for $a$ and $b$,  then (as seen in the appendix) we may use \eqref{circuit2} and \eqref{BCH} to calculate
\beq
\label{dyna}
\partial_t \mean{A'} = \tr{ -i [H_{loc}+ (\eta+i \xi)   A B, \rho_{ab} ] A' }\,,
\eeq 
\beq
\label{dynb}
\partial_t \mean{B'} = \tr{ -i [H_{loc}+ (\eta - i \xi)  A B, \rho_{ab} ] B' }\,,
\eeq 
where $H_{loc} = H_a + H_b+ \frac{\nu_a}{2} A^2 + \frac{\nu_b}{2}B^2$ contains only local terms and
\begin{eqnarray*}
\label{Hparams}
\nu_a &=&-i \Mean{ [x,\mc{S}^\dagger(x)]}_f\\ 
\nu_b &=&-i \Mean{ [p,\mc{S}^\dagger(p)]}_f \\
\eta &=& \Mean{ \frac{i}{2}([\mc{S}^\dagger(x),p]-[x,\mc{S}^\dagger(p)]) - 1 }_f \\
\xi &=& \Mean{ i - \frac{1}{2}([x,\mc{S}^\dagger(p)]+[\mc{S}^\dagger(x),p]) }_f \,.
\end{eqnarray*}
Thus in order for the average canonical variable dynamics to match the same classical Hamiltonian for both systems, we must have $\xi = 0$, or
\beq
\label{efest2}
\frac{1}{2}\Mean{ [x,\mc{S^\dagger}(p)] + [\mc{S^\dagger}(x),p] }_f = i\,.
\eeq

Given these physical assumptions we have a simple model for the emergence of non-local force laws by an intermediary. In the appendix we also calculate the associated dynamical equation \cite{Breuer2007}, 
\beq
\label{generator}
\mc{L}(\rho) = -i [\tilde H_\Sigma,\rho] + \mc{D}(\rho)\,.
\eeq
The first term reflects classical dynamics corresponding to the (level-shifted) Hamiltonian  
\beq
\label{Hamiltonian}
\tilde H_{\Sigma} = H_a + H_b + \frac{\nu_a}{2} A^2  + \frac{\nu_b}{2} B^2 + \eta AB\,.
\eeq
Alone it would produce a reversible evolution of $a$ and $b$, in contrast to the dissipative part of the generator, $\mc{D}$, which represents irreversible information exchange with the environment. Written explicitly the dissipator is
\begin{eqnarray}
\label{dissipator}
\mc{D}(\rho_{ab}) =& -\frac{1}{4}\l( Y_{xx}[A,[A,\rho_{ab}]] +Y_{pp} [B,[B,\rho_{ab}]] \r.\\
& \l. +2 Y_{xp}[A,[B,\rho_{ab}]] \r)\,, \nonumber
\end{eqnarray}
where the $Y$ parameters are determined by the screen's action on the FC, 
\begin{equation}
\label{params}
\begin{split}
Y_{xx} = {}& 2\Mean{ \mc{S}^\dagger(x^2) + x^2 - \{ x , \mc{S}^\dagger(x) \}   }_f  \\
Y_{pp} = {}& 2\Mean{ \mc{S}^\dagger(p^2) + p^2 -  \{p,\mc{S}^\dagger(p)\}}_f   \\
Y_{xp} = {} & \l\langle \mc{S}^\dagger( \{x,p\}) + \{x,p\}  - \{x,\mc{S}^\dagger(p)\} \r.  \\
{} &\l. - \{\mc{S}^\dagger(x),p\}   \r\rangle_f\,.
\end{split}
\end{equation}
In other contexts, the terms in $\mc{D}$ describe dynamics of systems undergoing weak measurements of operators $A$ and $B$, in which the measurement outcomes are being ignored\cite{Caves1987,Jacobs2006}. It is then natural to expect that classical behavior, by our definition, comes about when the measurement is too strong to allow for entanglement to develop. 

To see this is indeed the case in a specific context, we now assume that $a$ and $b$ correspond to single harmonic oscillators of unit mass and frequency. We take a linear interaction term $\eta A B = g x_a x_b$, where $x_s$ ($s = a,b$) is the position operator with canonical conjugate, $p_s$. Defining the vector $M = [x_a,p_a,x_b,p_b]^T$, we consider their covariance matrix, 
\beq
\label{gamma}
\gamma_{ij} = \tr{\rho_{ab}\, (M_i M_j + M_j M_i)} - 2 \tr{\rho_{ab}\, M_i}\tr{\rho_{ab}\, M_j}\,.
\eeq
A consequence of the positivity of the density matrix $\rho_{ab}$ is the Heisenberg uncertainty principle \cite{Kennard1927,Weyl1928,Robertson1929}, as implied by the operator inequality \cite{Lindblad2000,Holevo2011,Eisert2003}
\beq
\label{hberg}
\gamma + i \Delta_2 \geq 0\,,
\eeq
which means that the (complex valued) matrix $\gamma + i \Delta_2 $ has no negative eigenvalues. Here $\Delta_n$ represents the $n$-mode symplectic matrix, 
\beq
\Delta_n = \bigoplus_{i = 1}^n \l( \begin{array}{ccc} 0 & 1\\ -1& 0 \end{array} \r)_i\,.
\eeq

This admittedly complex notation allows us to quantify when classicality emerges, and it leads naturally to a local test of classical behavior.  Specifically,

\textbf{Lemma:} For the class of 2-mode Gaussian states  (i.e., states with vanishing cumulant tensors of higher order than $\gamma$)\cite{Caruso2008,Holevo2012} and any time $t>0$, evolution is classical -- i.e., entanglement cannot develop --  if and only if 
\beq
\label{cond}
Y_f - 2 i g \Delta_1 \geq 0\,,
\eeq
where $Y_f$ is the $2\times 2$ symmetric matrix composed of the terms in \eqref{params}. 

\textbf{Proof:} We note that a 2-mode Gaussian state is separable if and only if \cite{Simon2000}
\beq
\label{ppt}
\tilde  \gamma + i \Delta_2 \geq 0\,,
\eeq
where  $\tilde \gamma$ is obtained from $\gamma$ by setting $p_b\rightarrow -p_b$. In other words, 2-mode Gaussian states are entangled if and only if their partial time reverse does {\it not} satisfy the uncertainty principle. To see that this remains true under condition \eqref{cond}, we use \eqref{generator} to calculate the equation of motion of $\gamma$,
\beq
\label{dgamma}
\dot \gamma = x^T \gamma + \gamma x + y\,,
\eeq
where the matrices $x$ and $y$ are 
\beq
\label{dgammaparams}
x = - H \Delta_2 \quad y =  -\Delta_2 \chi Y_f \chi^T \Delta_2\,.
\eeq
The symmetric matrix $H$ parametrizes the Hamiltonian, $\frac{1}{2}\sum_{i j} H_{i j} M_i M_j = H_{\Sigma}$ and $Y$ defines the parameters in $\mc{D}$. The $4\times 2$ matrix $\chi$, with nonzero entry $1$ at $(i,j) = (1,1)$ and $(3,2)$, defines the correspondence between the system operators and force carrier quadratures $x$ and $p$, as set in the underlying propagator \eqref{circuit}. A straightforward calculation (in the appendix) then shows that, if \eqref{cond} is true, then for all $t>0$, a separable Gaussian state of $a$ and $b$ is always mapped to a separable Gaussian state under $e^{t \mc{L}}$. Conversely, if \eqref{cond} is not true, then there exists a $t>0$ such that the product state $\rho_{ab} = \ketbrad{0}\otimes\ketbrad{0}$ becomes entangled\footnote{In fact we may make a stronger statement about \eqref{cond}. If it is true, then {\it any} separable state is mapped to a state with no distillable entanglement \cite{Werner2001}. Conversely, if it is not true then there is a $t>0$ such that the ground state gains non-zero distillable entanglement. }. 

As we have verified in this simple example, classicality in interacting systems emerges as a consequence of decoherence due to environmental interactions, whose influence is represented by the dissipator $\mc{D}$. Although the FC is the intermediary allowing for communication between $a$ and $b$, it can also be viewed as a probe \cite{Braginsky1992} used in the weak measurement of these systems through the action of the screen, $\mc{S}$. As a consequence of the measurement of their positions, noise is introduced\cite{Heisenberg1927,arthurs1965,Busch2013} into their momenta that cannot be accounted for by the redistribution of noise due to Hamiltonian evolution in phase space. More concretely, if we define the ``reversible'' part of the covariance matrix as
\beq
\label{gammar}
\gamma_r(t) = e^{x^T t} \gamma(0) e^{x t}\,,
\eeq
which evolves according to \eqref{dgamma} with $y = 0$, then we would expect the momentum part of $(\gamma(t) -\gamma_r(t))$ to increase at a rate directly related to the measurement strength. This leads to a direct verification of classicality, which is the central result of this work. For clarity, we include all relevant units of $\hbar$, $m$, and $\omega$.

\textbf{Theorem}: If $a$ and $b$ are interacting in a classical way, then the excess rate of change of noise in $p_a$ and $p_b$ should exceed two times their coupling strength. That is, at all times $t$,
\beq
\label{nrate}
\partial_t \l(\mbox{Var}^{(e)}(p_a) + \mbox{Var}^{(e)}(p_b) \r)
\geq 2 |g| \hbar m \omega\,,
\eeq
where
\beq
\mbox{Var}^{(e)}(O) = \tr{(\rho_{ab}(t)-\rho_{ab}^{(r)}(t))\, O^2}\,.
\eeq 
The state $\rho_{ab}^{(r)}$ agrees with $\rho_{ab}$ at $t = 0$, and follows the reversible dynamics 
\beq
\partial_t \rho_{ab}^{(r)}(t) = -i[H_{local} - g m \omega x_a x_b,\rho^{(r)}(t)]\,.
\eeq
where $H_{local}$ represents all local terms in $H_{\Sigma}$.

\textbf{Proof}:
To see this, notice that for any (possibly non-Gaussian) state of $a$ and $b$, the covariance matrix $\gamma$ satisfies the equation of motion \eqref{dgamma}. The sum of variances above is then equal to $\frac{1}{2} z^\dagger (\gamma-\gamma_r) z $, where $z = [0,1, 0,i]^T$. Taking the time derivative produces the left hand side of Equation \eqref{nrate}, which -- at time $t = 0$ -- is just $\frac{1}{2}z^\dagger y z$. Hence, at $t = 0$, the inequality is a direct consequence of \eqref{dgammaparams} and \eqref{cond} (or its transpose, if $g <0$), which is equivalent to the classicality of the interaction. Since this statement is true for all covariance matrices $\gamma$, it must hold for $t>0$ as well.

Although our result is very specific, it can be extended to cases where $\mc{L}$ is no longer a Gaussian generator. Observe that since commutators involving canonical variables correspond to derivatives in their conjugates, the dynamical equation \eqref{dgamma} is unchanged if we add any terms to $\mc{L}$ involving such commutators of third or higher order. Hence our claims also apply to any generator $\mc{L}'$ of the form 
\beq
\mc{L}' = \mc{L} + \mc{G}\,,
\eeq
where $\mc{L}$ is defined as in \eqref{generator} and $\mc{G}$ contains only terms of the form $[M_{i_1},[M_{i_2},[M_{i_3},...]]]$ (with at least three nested commutators). This analysis leads us to speculate at a more general principle: if classical behavior is equivalent to lack of entanglement, it must be due to decoherence introduced by an effective measurement. Since measurement in one quadrature necessarily introduces noise into its canonical conjugate, classical behavior should always contain this signature of noise.

We conclude with a proposed experiment to demonstrate these concepts, in the context of the gravitational force, which provides rigor to the concepts elucidated in Ref.~\cite{page.1981.979--982}.  The general concept provided here -- a noise test of entanglement generation -- naturally extends to many other systems of relevance for efforts in quantum information science. 
\begin{figure}
\includegraphics[width=3.3in]{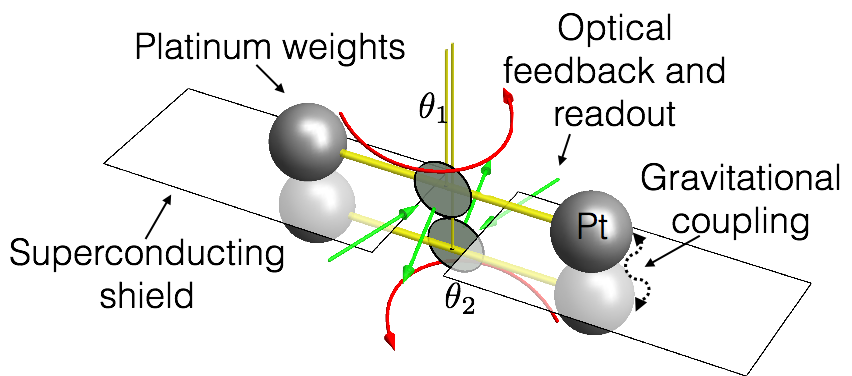}
\caption{
Schematic for a coupled torsional oscillator experiment, with many oscillator pairs.  Two adjacent, low frequency torsional oscillators have an observable gravitational coupling.  Measurement of the resonant power spectral noise density of the angular variables $\theta_1, \theta_2$ via a nearby optical waveguide detector should only be limited by temperature; classical forces would lead to excess noise beyond the thermal background, proportional to the gravitational coupling.
\label{f:expt}
}
\end{figure}
Our system consists of a pair of torsional oscillator attached to high density dumbbells, with mass $M$, radius $r$, and distance $R$ from their axis of rotation (see Fig.~2). These are arranged symmetrically at their equilibrium positions, so a relative angular displacement $\theta$ produces a gravitational restoring force proportional to $\theta R \frac{G M^2}{r^3}$. Expanding to leading order in their angular displacements, we get an interacting Hamiltonian
\beq
H_{grav} = \frac{L_a^2 + L_b^2}{2 I}  + \frac{1}{2}I \omega(\omega+g) (\theta_a^2 +\theta_b^2) -  I \omega g\theta_a \theta_b\,,
\eeq
where $I\sim 2 M R^2$ is the moment of inertia of each dumbbell, $\omega$ the bare spring resonant frequency, and $L_s$ the angular momentum. Up to a geometric factor of order $\sim 1$, the gravitational coupling is $ g \sim G n/\omega $, where $G$ is Newton's constant and $n$ the mass density of the spheres. 

The parameters in $H_{grav}$ can be determined experimentally by observing the average evolution of the canonical variables $\theta_s$ and $L_s$. Realistically, we expect these values to dissipate over time due to local interactions with a thermal environment, but the associated dissipation rate $\kappa$ may also be determined by tracking the canonical variables. Incorporating this decay into the expected evolution of $L_a^2$ and $L_b^2$, we have that for a classical interaction,
\beq
\label{Gineq}
(\partial_t + \kappa )\l(\mbox{Var}^{(e)}(L_a) + \mbox{Var}^{(e)}(L_b) \r) \geq 2 g \hbar I \omega\,.
\eeq

Naturally, our proposed test of classicality is feasible only if we can measure the rate of change of $L_s$ and $L_s^2$ over a reasonable time scale. Towards this end, developing independent measurements of the thermal noise and the excess noise we are testing for becomes extremely challenging in the small $\kappa$ limit, due to thermal correlations in the variance.  However, using cold damping techniques \cite{Milatz1953,Mancini1998,Cohadon1999,LIGO2009,Lee2010}, we can achieve a reasonable integration time.  To understand the cold-damping approach as applied here, one conceives of a single `shot' of the experiment as follows.  With the oscillators clamped at some initial time, release the clamps, and allow for oscillation due to all sources of noise for a time $\tau \geq 1/g$.  Then use measurement and feedback on a time scale $\leq \tau$ to simultaneously return the oscillators to the clamped state and to estimate the overall amplitude of oscillations induced by the noise during that time. The feedback cooling of the system provides statistically independent samples over time scales on the order of $1/g$, as opposed to waiting for the system to reach its steady state at a rate $\kappa \ll g$. 

For a single shot integration time $\tau$, the scale of noise in $L^2$ due to thermal fluctuations is $\hbar I \omega (\bar n \kappa \tau)$, where $\bar n \approx \frac{k_B T}{\hbar \omega} \gg 1$ is the steady state occupation of phonons at high temperatures. Comparing this with the excess noise signal $\tau g \hbar I \omega$ (from Equation \eqref{Gineq}), we have a signal-to-noise after $N_s  \approx T_{int} / (2 \tau)$ shots of
\beq
\frac{S}{N} \approx \frac{g}{\bar n \kappa} \sqrt{\frac{T_{int}}{2 \tau}}
\eeq
We see that the total integration time $T_{int}$ for a $5 \sigma$ result scales as
\beq
T_{int} \leq 50 \tau \left(\frac{ \bar n \kappa }{  g} \right)^2 
\approx \frac{50}{g} \left(\frac{k_b T}{\hbar g Q} \right)^2
\eeq
For a platinum dumbbell ($n = 22 $ g$/\mbox{cm}^3$), a frequency $\omega = 1 \mbox{ mHz}$ oscillator common in small-scale tests of Newton's laws, a mechanical $Q \sim 10^9$, and assumed temperature $10$ mK in a dilution fridge, this produces a coupling $g \sim 0.23$ mHz and a $5\sigma$ integration time on the order of a few thousand years.  Fortunately, each `shot' of the experiment would only be a few thousand seconds, which would enable measurement of many parallel devices to achieve the desired precision goal in substantially shorter real world time.  In particular, the ability to mass fabricate high quality oscillators using modern techniques from chip-scale opto-mechanics suggest a potential path to a low cost system with thousands of devices.

We recognize the difficulty of the proposed experiment -- these rates correspond to observation of one excess phonon per few hours against a large thermal background -- as well as concede that this tabletop test is far from loop-hole free.  Crucial challenges to be addressed in any serious attempt to bound this noise inequality will have to work against both systematic effects, such as temperature variations in the laboratory setting and screening of electromagnetic coupling between torsional oscillators, but also against unresolved challenges regarding quantization of collective degrees of freedom associated with macroscopic masses.  However, 
ultra-high Q, low frequency torsional oscillators approaching the necessary requirements and thermally-limited performance have been demonstrated in the laboratory~\cite{UW,bantel.2000.233--242}.  
Furthermore, such an observation showing the lack of excess noise would both validate the perturbative model of gravitons and its noise properties~\cite{calzetta.1994.6636--6655}, provide stringent constraint on semiclassical gravity theories and information non-conserving theories, and provide insight into the fundamental question of whether gravity can convey quantum information.

\acknowledgements{We thank Gerard Milburn, Anders Sorensen, Elizabeth Goldschmidt, Edward Baker III, John Gough, John Preskill, and Bei-Lok Hu for helpful discussions and insights, and Emily Edwards for visualization.  This work was supported by the NSF-supported Physical Frontier Center at the JQI.}

\section*{Appendix}
\subsection{Calculation of the generator and physical considerations}

Since we only consider observables of the form $O_{ab}\otimes \mbb{I}_f$ in the reduced dynamics of $a$ and $b$, the calculation is made simpler in the Heisenberg picture. Hence we will consider the adjoint circuit superoperator
\begin{equation}
\label{circuit3}
\mc{V}_t^\dagger =  \mc{C}_{-A x \sqrt{t}}\, \mc{C}_{-B p \sqrt{t}}\,  \mc{S}^\dagger \, \mc{C}_{A x \sqrt{t}} \, \mc{C}_{B p \sqrt{t}}\,\mc{C}_{-(H_a+H_b)t}\, ,
\end{equation}
where $\mc{S}^\dagger$ is defined by the relation $\tr{ O \mc{S}^\dagger(O')} \equiv \tr{\mc{S}(O) O' }$. The reduced circuit superoperator, acting on operators on $\mc{H}_{a b}$, is then
\begin{equation}
\label{circuit4}
(\mc{V}_t^{red})^\dagger(O_{ab}) = \mbox{tr}_{f}\{\mbb{I}_{ab}\otimes \rho_f \,\mc{V}_t^\dagger(O_{ab}\otimes \mbb{I}_f) \}
\end{equation}

To show when the limit \eqref{trotter} exists, we use the Baker-Campbell-Hausdorff relation to expand each term of \eqref{circuit3} to first order in $\sqrt{t}$. For example, we may expand $\mc{C}_{B p \sqrt{t}}(O) = e^{-i B p \sqrt{t}} O e^{i B p \sqrt{t}} $ as
\begin{equation}
\label{BCHexpand}
\mc{C}_{B p \sqrt{t}} = \mbox{Id} - i \sqrt{t} [B p, \, \cdot \, ] - \frac{t}{2}  [B p,  [B p, \, \cdot \, ]] + O(t^{3/2})\,,
\end{equation}
where $\mbox{Id}(O) = O$ is the identity super-operator, and $[B p, \, \cdot \, ]O = [B p, O]$ is the commutation super-operator. Adding all terms of $\mc{V}_t^\dagger$ of order $\sqrt{t}$ gives
\begin{equation}
-i \sqrt{t}  [B , \, \cdot \, ]\otimes\l(\mc{S}^\dagger(p)-p\r) - i \sqrt{t}  [A , \, \cdot \, ]\otimes\l(\mc{S}^\dagger(x)-x\r) \,,
\end{equation}
where we have pulled out $x$ and $p$ from the commutators since we are only acting on terms of the form $O_{ab}\otimes \mbb{I}_f$. Taking the trace in \eqref{circuit4}, we have that the $\sqrt{t}$ terms of $\mc{V}^\dagger_t$ vanish exactly when equation \eqref{efest1} holds. 

To calculate the dynamics induced by the reduced circuit \eqref{circuit2}, we continue using Baker-Campbell-Hausdorff (as in Equation \eqref{BCHexpand}) and expand each superoperator in $\mc{V}_t^\dagger$ to order $t$.  The adjoint generator $\mc{L}^{\dagger}$, from which we can compute $\mc{L}$, is obtained by tracing out the FC,
\begin{equation}
\label{Ldag}
\mc{L}^\dagger(O_{ab}) = \mbox{tr}_f\{\mbb{I}\otimes \rho_f \,\l( \partial_t \mc{V}^\dagger_t\r)|_{t = 0} (O_{ab}\otimes \mbb{I}) \}\,.
\end{equation}
There are multiple sources of order $t$ terms in $\mc{V}_t^\dagger$. The first is just the commutator
\begin{equation}
\label{local}
i[H_a + H_b, \, \cdot \,]
\end{equation}
obtained from the local unitary $\mc{C}_{-(H_a + H_b)t}$. Second, we have products of order $\sqrt{t}$ arising from commutators of different FC-system interactions (e.g. $\mc{C}_{\sqrt{t} A x}$ and $\mc{C}_{\sqrt{t} B p}$), which -- after accounting for the action of the screen -- produce
\begin{widetext}
\begin{eqnarray}
\label{cross}
\begin{split}
i[A B, \, \cdot \,]\otimes\l(  \frac{i}{2}([\mc{S}^\dagger(x),p]-[x,\mc{S}^\dagger(p)]) + 1 \r)  
 & + \l(A \, \cdot \, B - B \, \cdot \, A \r)\otimes \l( i - \frac{1}{2}([x,\mc{S}^\dagger(p)]+[\mc{S}^\dagger(x),p]) \r) \\
&-\frac{1}{2}[A,[B,\, \cdot \,]]\otimes \l( \{x,p \} + \mc{S}^\dagger(\{x,p \}) - \{\mc{S}^\dagger(x),p \}-\{x,\mc{S}^\dagger(p) \} \r) \,,
\end{split}
\end{eqnarray}
\end{widetext}
where we have used $[x,p] = i$. This calculation also uses the fact that all inputs to $\mc{V}_t^\dagger $ are of the form $O_{ab}\otimes\mbb{I}_f$, allowing us  to remove terms from the commutators with the Jacobi identity $[A B, C] = A[B,C] + [A,C]B$. 

Finally, we have products of order $\sqrt{t}$ having strictly $A x$ or $Bp$ commutators (e.g., from $\mc{C}_{\sqrt{t} A x}$ and $\mc{C}_{- \sqrt{t} Ax }$) , as well as individual order $t$ terms involving double commutators from a single interaction (e.g. the third term of \eqref{BCHexpand}). This gives
\begin{eqnarray}
\label{self}
\begin{split}
&i[\frac{1}{2}A^2, \, \cdot \,]\otimes\l( -i [x,\mc{S}[x]] \r) & \\
&-\frac{1}{2}[A,[A,\, \cdot \,]]\otimes \l( x^2 + \mc{S}(x^2) - \{x,\mc{S}(x)\} \r)& \,.
\end{split}
\end{eqnarray}
and an analogous contribution from the $B$ terms, obtained by substituting $x\rightarrow p$ and $A \rightarrow B$ above. 

Adding all terms together and taking the trace over the FC, we have
\begin{eqnarray}
\label{Ldag2}
\begin{split}
\mc{L}^\dagger (O_{ab}) = & i[H_a + H_b + \frac{\nu_a}{2}A^2 +  \frac{\nu_b}{2}B^2 + \eta AB, O_{ab}] \\
&  + \xi \l(A \, O_{ab} \, B - B \, O_{ab} \, A \r)  \\
& \mc{D}(O_{ab}) \,
\end{split}
\end{eqnarray}
The level shift factor $\nu_a$ comes from the first term of \eqref{self} and $\nu_b$ from its $Bp$ analogue (these are defined exactly as in \eqref{Hparams}.) The dissipator $\mc{D}$, defined in \eqref{dissipator} and \eqref{params}, comes from the final terms of Equations \eqref{self}, its $B p$ analogue, and \eqref{cross}. The remaining terms, coming from the first and second terms of \eqref{cross}, are those proportional to 
\begin{equation}
\eta =  \Mean{ \frac{i}{2}([\mc{S}^\dagger(x),p]-[x,\mc{S}^\dagger(p)]) - 1 }_f
\end{equation}
and 
\begin{equation}
 \xi =  \Mean{ i - \frac{1}{2}([x,\mc{S}^\dagger(p)]+[\mc{S}^\dagger(x),p]) }_f \,.
\end{equation}  

To justify constraint \eqref{efest2} and complete the calculation of $\mc{L}^\dagger$, we assume that $A$ and $B$ are canonical operators of $a$ and $b$, respectively. Then for any other such canonical operators, $A'$ and $B'$, we immediately have that $\mc{D}(A') = \mc{D}(B') = 0$ (since $[A,A']$, $[B, A']$, $[A,B']$ and $[B,B']$ are each a number.) The Heisenberg equations of motion are then
\begin{equation}
\mc{L}^\dagger(A') = i [ H_{loc}+ (\eta -i\xi)   A B, A'] \,,
\end{equation} 
\begin{equation}
\mc{L}^\dagger(B') = i [ H_{loc}+ (\eta+ i\xi)   A B, B']\,,
\end{equation}
where we have set $H_{loc} = H_a + H_b + \frac{\nu_a}{2} A^2+  \frac{\nu_b}{2} B^2$. We now impose Ehrenfest's theorem, which implies that all canonical variables evolve according to the (same) classical Hamiltonian, and therefore must assume that $\xi = 0$. This is equivalent to the constraint \eqref{efest2}, and by substitution implies that $\eta = i\Mean{[\mc{S}^\dagger(x),p]}_f$, in agreement with the penultimate equation of \eqref{Hparams}. Applying these constraints to \eqref{Ldag2}, we have
\begin{equation}
\mc{L}^\dagger (O_{ab}) =  i[\tilde H_\Sigma , O_{ab}] + \mc{D}(O_{ab}) \,
\end{equation}
where $\tilde H_\Sigma$ is defined in \eqref{Hamiltonian}. In order to obtain $\mc{L}$ from $\mc{L}^\dagger$, we note the cyclic property of the trace ($\tr{O\, O'} = \tr{O'\, O}$), from which it is easy to check that $( i[\tilde H_\Sigma , \, \cdot \,])^\dagger = -i[\tilde H_\Sigma , \, \cdot \,]$ and $\mc{D} = \mc{D}^\dagger$. 

\subsection{Gaussian dynamics and entanglement generation}
Equation \eqref{dgamma} may be derived using the generator $\partial_t \rho_{ab} = \mc{L}(\rho_{ab})$ in \eqref{generator} and the definition of $\gamma$ in \eqref{gamma}, noting that $i (\Delta_2)_{i j} = [M_i,M_j]$. To see that condition \eqref{cond} implies no entanglement generation in Gaussian states, we first note that since it only involves second order commutators of canonical variables, the generator $\mc{L}$ necessarily maps Gaussian states to Gaussian states\cite{Heinosaari2010}. The differential equation \eqref{dgamma} may be solved directly to give
\begin{equation}
\label{gammat}
\gamma(t) = Y_t + X_t^T \gamma(0) X_t\,,
\end{equation}
where $X_t = e^{x t}$ and $Y_t = \int_0^t  X_{t-s}^T\, y\, X_{t-s} \mbox{d}s$. A 2-mode Gaussian state is separable if and only if its covariance matrix satisfies \eqref{ppt} \cite{Simon2000}. Hence it suffices to show that if any $\gamma(0)$ satisfies this relation, $\gamma(t)$ does so as well for sufficiently small $t>0$. 

Given that $\tilde \gamma(0) = K \gamma(0) K$, with $K =\mbox{diag}([1,1,1,-1])$, we have that \eqref{ppt} is equivalent to
\begin{equation}
\label{ppt2}
\gamma(0) \geq - i K \Delta_2 K \equiv -i \tilde \Delta_2\,.
\end{equation}
Therefore what we must show is that if \eqref{ppt2} is true and there is some vector $z$ such that $z^\dagger\l(\gamma + i \tilde \Delta_2\r) z = 0$, then $z^\dagger \gamma'(0)z \geq 0$, which implies that \eqref{ppt2} holds for small $t>0$. To do this, we use \eqref{gammat} and \eqref{ppt2} to compute
\begin{eqnarray}
\label{Fineq}
\begin{split}
\gamma(t) + i \tilde \Delta_2 &= Y_t + X_t^T \gamma(0) X_t + i \tilde \Delta_2\\
& \geq  Y_t - X_t^T i \tilde \Delta_2 X_t + i \tilde \Delta_2 \\
& \equiv  F(t)\,.
\end{split}
\end{eqnarray}
Since $F(0) = 0$, it suffices to show that $F'(0)\geq 0$, for if $z^\dagger (\gamma(0) +i \tilde \Delta_2)z = 0$, then $z^\dagger \gamma'(0) z<0$ would produce a contradiction with \eqref{Fineq}.  Using \eqref{dgammaparams}, a straightforward calculation shows that
\begin{eqnarray}
F'(0) &=& y - i x^T \tilde \Delta_2 - i \tilde \Delta_2 x\\
& = &  - \Delta_2 \chi Y_f \chi^T \Delta_2 - i \Delta_2 H \tilde \Delta_2 + i \tilde \Delta_2 H  \Delta_2\\
& =& (i \Delta_2) \chi (Y_f - 2 i g  \Delta_1) \chi^T (i \Delta_2)\,.
\end{eqnarray}
Since $i\Delta_2$ is a Hermitian matrix, $F'(0)\geq 0$ follows immediately from \eqref{cond}. 

To prove the converse statement, we assume that \eqref{cond} is not true, so that $(Y_f - 2 i g  \Delta_1)$ has a negative eigenvalue. Assuming  the systems each start in their ground state, the initial covariance matrix is simply the identity, $\gamma(0) = I$. Expanding about $t = 0$, we have
\begin{equation}
\gamma(t) +i \tilde \Delta_2 = I + i \tilde \Delta_2 + t(y + x^T + x) + O(t^2)\,.
\end{equation} 
Since \eqref{cond} is not true, there exists a complex valued vector $z_f = [z_1,z_2]^T$ such that $z_f^\dagger (Y_f - 2 i g  \Delta_1) z_f <0$. Set $z_{ab} = [-z_1, i z_1, z_2, i z_2 ]^T$ so that $\chi^T i \Delta_2 z_{ab}= z_f$ and $i \tilde \Delta_2 z_{a b} = - z_{a b}$. The zero order term of $z^\dagger(\gamma(t) + i \tilde \Delta_2) z $ then vanishes, while the first order term is 
\begin{equation}
z^\dagger (y + x^T + x) z = z^\dagger ( y - i x^T \tilde \Delta_2 - i \tilde \Delta_2 x )z = z_f^\dagger (Y_f - 2 i g  \Delta_1) z_f\,,
\end{equation}
which is strictly negative. This shows that for small $t$ the matrix $\gamma(t)$ does not satisfy \eqref{ppt}, so that the ground state becomes a state with negative partial transpose, and is therefore entangled.


\begin{thebibliography}{46}
\expandafter\ifx\csname natexlab\endcsname\relax\def\natexlab#1{#1}\fi
\expandafter\ifx\csname bibnamefont\endcsname\relax
  \def\bibnamefont#1{#1}\fi
\expandafter\ifx\csname bibfnamefont\endcsname\relax
  \def\bibfnamefont#1{#1}\fi
\expandafter\ifx\csname citenamefont\endcsname\relax
  \def\citenamefont#1{#1}\fi
\expandafter\ifx\csname url\endcsname\relax
  \def\url#1{\texttt{#1}}\fi
\expandafter\ifx\csname urlprefix\endcsname\relax\def\urlprefix{URL }\fi
\providecommand{\bibinfo}[2]{#2}
\providecommand{\eprint}[2][]{\url{#2}}

\bibitem[{\citenamefont{Weinberg}(1995)}]{Weinberg1995}
\bibinfo{author}{\bibfnamefont{S.}~\bibnamefont{Weinberg}},
  \emph{\bibinfo{title}{The Quantum Theory of Fields}}
  (\bibinfo{publisher}{Cambridge University Press}, \bibinfo{year}{1995}), ISBN
  \bibinfo{isbn}{9780521585552},
  \urlprefix\url{http://books.google.com/books?id=bNDFnQEACAAJ}.

\bibitem[{\citenamefont{Srednicki}(2007)}]{Srednicki2007}
\bibinfo{author}{\bibfnamefont{M.}~\bibnamefont{Srednicki}},
  \emph{\bibinfo{title}{Quantum Field Theory}} (\bibinfo{publisher}{Cambridge
  University Press}, \bibinfo{year}{2007}).

\bibitem[{\citenamefont{Zee}(2010)}]{Zee2010}
\bibinfo{author}{\bibfnamefont{A.}~\bibnamefont{Zee}},
  \emph{\bibinfo{title}{Quantum Field Theory in a Nutshell: (Second Edition)}},
  In a Nutshell (\bibinfo{publisher}{Princeton University Press},
  \bibinfo{year}{2010}), ISBN \bibinfo{isbn}{9781400835324},
  \urlprefix\url{http://books.google.com/books?id=n8Mmbjtco78C}.

\bibitem[{\citenamefont{Bell}(1964)}]{Bell1964}
\bibinfo{author}{\bibfnamefont{J.~S.} \bibnamefont{Bell}},
  \bibinfo{journal}{Physics} \textbf{\bibinfo{volume}{1}}, \bibinfo{pages}{195}
  (\bibinfo{year}{1964}).

\bibitem[{\citenamefont{Aspect et~al.}(1981)\citenamefont{Aspect, Grangier, and
  Roger}}]{Aspect1981}
\bibinfo{author}{\bibfnamefont{A.}~\bibnamefont{Aspect}},
  \bibinfo{author}{\bibfnamefont{P.}~\bibnamefont{Grangier}}, \bibnamefont{and}
  \bibinfo{author}{\bibfnamefont{G.}~\bibnamefont{Roger}},
  \bibinfo{journal}{Phys. Rev. Lett.} \textbf{\bibinfo{volume}{47}},
  \bibinfo{pages}{460} (\bibinfo{year}{1981}),
  \urlprefix\url{http://link.aps.org/doi/10.1103/PhysRevLett.47.460}.

\bibitem[{\citenamefont{Clauser et~al.}(1969)\citenamefont{Clauser, Horne,
  Shimony, and Holt}}]{Clauser1969}
\bibinfo{author}{\bibfnamefont{J.~F.} \bibnamefont{Clauser}},
  \bibinfo{author}{\bibfnamefont{M.~A.} \bibnamefont{Horne}},
  \bibinfo{author}{\bibfnamefont{A.}~\bibnamefont{Shimony}}, \bibnamefont{and}
  \bibinfo{author}{\bibfnamefont{R.~A.} \bibnamefont{Holt}},
  \bibinfo{journal}{Phys. Rev. Lett.} \textbf{\bibinfo{volume}{23}},
  \bibinfo{pages}{880} (\bibinfo{year}{1969}),
  \urlprefix\url{http://link.aps.org/doi/10.1103/PhysRevLett.23.880}.

\bibitem[{\citenamefont{Caldeira and Leggett}(1983)}]{Caldeira1983}
\bibinfo{author}{\bibfnamefont{A.}~\bibnamefont{Caldeira}} \bibnamefont{and}
  \bibinfo{author}{\bibfnamefont{A.}~\bibnamefont{Leggett}},
  \bibinfo{journal}{Physica A: Statistical Mechanics and its Applications}
  \textbf{\bibinfo{volume}{121}}, \bibinfo{pages}{587 } (\bibinfo{year}{1983}),
  ISSN \bibinfo{issn}{0378-4371},
  \urlprefix\url{http://www.sciencedirect.com/science/article/pii/0378437183900134}.

\bibitem[{\citenamefont{Joos and Zeh}(1985)}]{Joos1985}
\bibinfo{author}{\bibfnamefont{E.}~\bibnamefont{Joos}} \bibnamefont{and}
  \bibinfo{author}{\bibfnamefont{H.}~\bibnamefont{Zeh}},
  \bibinfo{journal}{Zeitschrift f\"ur Physik B Condensed Matter}
  \textbf{\bibinfo{volume}{59}}, \bibinfo{pages}{223} (\bibinfo{year}{1985}),
  ISSN \bibinfo{issn}{0722-3277},
  \urlprefix\url{http://dx.doi.org/10.1007/BF01725541}.

\bibitem[{\citenamefont{Di{\~A}³si}(1987)}]{Diosi1987}
\bibinfo{author}{\bibfnamefont{L.}~\bibnamefont{Di{\~A}³si}},
  \bibinfo{journal}{Physics Letters A} \textbf{\bibinfo{volume}{120}},
  \bibinfo{pages}{377 } (\bibinfo{year}{1987}), ISSN \bibinfo{issn}{0375-9601},
  \urlprefix\url{http://www.sciencedirect.com/science/article/pii/0375960187906815}.

\bibitem[{\citenamefont{Di\'osi}(1989)}]{Diosi1989}
\bibinfo{author}{\bibfnamefont{L.}~\bibnamefont{Di\'osi}},
  \bibinfo{journal}{Phys. Rev. A} \textbf{\bibinfo{volume}{40}},
  \bibinfo{pages}{1165} (\bibinfo{year}{1989}),
  \urlprefix\url{http://link.aps.org/doi/10.1103/PhysRevA.40.1165}.

\bibitem[{\citenamefont{van Wezel and Oosterkamp}(2012)}]{vanWezel2012}
\bibinfo{author}{\bibfnamefont{J.}~\bibnamefont{van Wezel}} \bibnamefont{and}
  \bibinfo{author}{\bibfnamefont{T.~H.} \bibnamefont{Oosterkamp}},
  \bibinfo{journal}{Proceedings of the Royal Society A: Mathematical, Physical
  and Engineering Science} \textbf{\bibinfo{volume}{468}}, \bibinfo{pages}{35}
  (\bibinfo{year}{2012}),
  \urlprefix\url{http://rspa.royalsocietypublishing.org/content/468/2137/35.abstract}.

\bibitem[{\citenamefont{Gell-Mann and Hartle}(1993)}]{Gell-Mann1993}
\bibinfo{author}{\bibfnamefont{M.}~\bibnamefont{Gell-Mann}} \bibnamefont{and}
  \bibinfo{author}{\bibfnamefont{J.~B.} \bibnamefont{Hartle}},
  \bibinfo{journal}{Phys. Rev. D} \textbf{\bibinfo{volume}{47}},
  \bibinfo{pages}{3345} (\bibinfo{year}{1993}),
  \urlprefix\url{http://link.aps.org/doi/10.1103/PhysRevD.47.3345}.

\bibitem[{\citenamefont{Ehrenfest}(1927)}]{Ehrenfest1927}
\bibinfo{author}{\bibfnamefont{P.}~\bibnamefont{Ehrenfest}},
  \bibinfo{journal}{Zeitschrift f{\~A}¼r Physik}
  \textbf{\bibinfo{volume}{45}}, \bibinfo{pages}{455} (\bibinfo{year}{1927}),
  ISSN \bibinfo{issn}{0044-3328},
  \urlprefix\url{http://dx.doi.org/10.1007/BF01329203}.

\bibitem[{\citenamefont{Yang et~al.}(2013)\citenamefont{Yang, Miao, Lee, Helou,
  and Chen}}]{Yang2013}
\bibinfo{author}{\bibfnamefont{H.}~\bibnamefont{Yang}},
  \bibinfo{author}{\bibfnamefont{H.}~\bibnamefont{Miao}},
  \bibinfo{author}{\bibfnamefont{D.-S.} \bibnamefont{Lee}},
  \bibinfo{author}{\bibfnamefont{B.}~\bibnamefont{Helou}}, \bibnamefont{and}
  \bibinfo{author}{\bibfnamefont{Y.}~\bibnamefont{Chen}},
  \bibinfo{journal}{Phys. Rev. Lett.} \textbf{\bibinfo{volume}{110}},
  \bibinfo{pages}{170401} (\bibinfo{year}{2013}),
  \urlprefix\url{http://link.aps.org/doi/10.1103/PhysRevLett.110.170401}.

\bibitem[{\citenamefont{Milburn et~al.}(2000)\citenamefont{Milburn, Schneider,
  and James}}]{Milburn2000}
\bibinfo{author}{\bibfnamefont{G.}~\bibnamefont{Milburn}},
  \bibinfo{author}{\bibfnamefont{S.}~\bibnamefont{Schneider}},
  \bibnamefont{and} \bibinfo{author}{\bibfnamefont{D.}~\bibnamefont{James}},
  \bibinfo{journal}{Fortschritte der Physik} \textbf{\bibinfo{volume}{48}},
  \bibinfo{pages}{801} (\bibinfo{year}{2000}), ISSN \bibinfo{issn}{1521-3978},
  \urlprefix\url{http://dx.doi.org/10.1002/1521-3978(200009)48:9/11<801::AID-PROP801>3.0.CO;2-1}.

\bibitem[{\citenamefont{M\o{}lmer and S\o{}rensen}(1999)}]{Molmer1999}
\bibinfo{author}{\bibfnamefont{K.}~\bibnamefont{M\o{}lmer}} \bibnamefont{and}
  \bibinfo{author}{\bibfnamefont{A.}~\bibnamefont{S\o{}rensen}},
  \bibinfo{journal}{Phys. Rev. Lett.} \textbf{\bibinfo{volume}{82}},
  \bibinfo{pages}{1835} (\bibinfo{year}{1999}).

\bibitem[{\citenamefont{Garc\'\i{}a-Ripoll
  et~al.}(2005)\citenamefont{Garc\'\i{}a-Ripoll, Zoller, and
  Cirac}}]{Garcia-Ripoll2005}
\bibinfo{author}{\bibfnamefont{J.~J.} \bibnamefont{Garc\'\i{}a-Ripoll}},
  \bibinfo{author}{\bibfnamefont{P.}~\bibnamefont{Zoller}}, \bibnamefont{and}
  \bibinfo{author}{\bibfnamefont{J.~I.} \bibnamefont{Cirac}},
  \bibinfo{journal}{Phys. Rev. A} \textbf{\bibinfo{volume}{71}},
  \bibinfo{pages}{062309} (\bibinfo{year}{2005}).

\bibitem[{\citenamefont{Nielsen and Chuang}(2000)}]{Nielsen2000}
\bibinfo{author}{\bibfnamefont{M.~A.} \bibnamefont{Nielsen}} \bibnamefont{and}
  \bibinfo{author}{\bibfnamefont{I.~L.} \bibnamefont{Chuang}},
  \emph{\bibinfo{title}{Quantum Computation and Quantum Information}}
  (\bibinfo{publisher}{Cambridge University Press}, \bibinfo{year}{2000}).

\bibitem[{\citenamefont{Lindblad}(1976)}]{Lindblad1976}
\bibinfo{author}{\bibfnamefont{G.}~\bibnamefont{Lindblad}},
  \bibinfo{journal}{Communications in Mathematical Physics}
  \textbf{\bibinfo{volume}{48}}, \bibinfo{pages}{119} (\bibinfo{year}{1976}),
  ISSN \bibinfo{issn}{0010-3616},
  \urlprefix\url{http://dx.doi.org/10.1007/BF01608499}.

\bibitem[{\citenamefont{Breuer and Petruccione}(2007)}]{Breuer2007}
\bibinfo{author}{\bibfnamefont{H.}~\bibnamefont{Breuer}} \bibnamefont{and}
  \bibinfo{author}{\bibfnamefont{F.}~\bibnamefont{Petruccione}},
  \emph{\bibinfo{title}{The Theory of Open Quantum Systems}}
  (\bibinfo{publisher}{Clarendon Press}, \bibinfo{year}{2007}), ISBN
  \bibinfo{isbn}{9780199213900},
  \urlprefix\url{http://books.google.com/books?id=DkcJPwAACAAJ}.

\bibitem[{\citenamefont{Caves and Milburn}(1987)}]{Caves1987}
\bibinfo{author}{\bibfnamefont{C.~M.} \bibnamefont{Caves}} \bibnamefont{and}
  \bibinfo{author}{\bibfnamefont{G.~J.} \bibnamefont{Milburn}},
  \bibinfo{journal}{Phys. Rev. A} \textbf{\bibinfo{volume}{36}},
  \bibinfo{pages}{5543} (\bibinfo{year}{1987}),
  \urlprefix\url{http://link.aps.org/doi/10.1103/PhysRevA.36.5543}.

\bibitem[{\citenamefont{Jacobs and Steck}(2006)}]{Jacobs2006}
\bibinfo{author}{\bibfnamefont{K.}~\bibnamefont{Jacobs}} \bibnamefont{and}
  \bibinfo{author}{\bibfnamefont{D.~A.} \bibnamefont{Steck}},
  \bibinfo{journal}{Contemporary Physics} \textbf{\bibinfo{volume}{47}},
  \bibinfo{pages}{279} (\bibinfo{year}{2006}),
  \urlprefix\url{http://www.tandfonline.com/doi/abs/10.1080/00107510601101934}.

\bibitem[{\citenamefont{Kennard}(1927)}]{Kennard1927}
\bibinfo{author}{\bibfnamefont{E.}~\bibnamefont{Kennard}},
  \bibinfo{journal}{Zeitschrift f{\~A}¼r Physik}
  \textbf{\bibinfo{volume}{44}}, \bibinfo{pages}{326} (\bibinfo{year}{1927}),
  ISSN \bibinfo{issn}{0044-3328},
  \urlprefix\url{http://dx.doi.org/10.1007/BF01391200}.

\bibitem[{\citenamefont{Weyl}(1928)}]{Weyl1928}
\bibinfo{author}{\bibfnamefont{H.}~\bibnamefont{Weyl}},
  \emph{\bibinfo{title}{Gruppentheorie und Quantenmechanik. - Leipzig, Hirzel
  1928. VIII, 288 S.}} (\bibinfo{publisher}{Leipzig}, \bibinfo{year}{1928}),
  \urlprefix\url{http://books.google.com/books?id=-VReAAAAIAAJ}.

\bibitem[{\citenamefont{Robertson}(1929)}]{Robertson1929}
\bibinfo{author}{\bibfnamefont{H.~P.} \bibnamefont{Robertson}},
  \bibinfo{journal}{Phys. Rev.} \textbf{\bibinfo{volume}{34}},
  \bibinfo{pages}{163} (\bibinfo{year}{1929}),
  \urlprefix\url{http://link.aps.org/doi/10.1103/PhysRev.34.163}.

\bibitem[{\citenamefont{Lindblad}(2000)}]{Lindblad2000}
\bibinfo{author}{\bibfnamefont{G.}~\bibnamefont{Lindblad}},
  \bibinfo{journal}{Journal of Physics A: Mathematical and General}
  \textbf{\bibinfo{volume}{33}}, \bibinfo{pages}{5059} (\bibinfo{year}{2000}),
  \urlprefix\url{http://stacks.iop.org/0305-4470/33/i=28/a=310}.

\bibitem[{\citenamefont{Holevo}(2011)}]{Holevo2011}
\bibinfo{author}{\bibfnamefont{A.}~\bibnamefont{Holevo}},
  \emph{\bibinfo{title}{Probabilistic and Statistical Aspects of Quantum
  Theory}}, Monographs quaderni (\bibinfo{publisher}{Scuola Normale Superiore},
  \bibinfo{year}{2011}), ISBN \bibinfo{isbn}{9788876423789},
  \urlprefix\url{http://books.google.com/books?id=l7AIDhbWrTIC}.

\bibitem[{\citenamefont{Eisert and Plenio}(2003)}]{Eisert2003}
\bibinfo{author}{\bibfnamefont{J.}~\bibnamefont{Eisert}} \bibnamefont{and}
  \bibinfo{author}{\bibfnamefont{M.~B.} \bibnamefont{Plenio}},
  \bibinfo{journal}{International Journal of Quantum Information}
  \textbf{\bibinfo{volume}{01}}, \bibinfo{pages}{479} (\bibinfo{year}{2003}),
  \urlprefix\url{http://www.worldscientific.com/doi/abs/10.1142/S0219749903000371}.

\bibitem[{\citenamefont{Caruso et~al.}(2008)\citenamefont{Caruso, Eisert,
  Giovannetti, and Holevo}}]{Caruso2008}
\bibinfo{author}{\bibfnamefont{F.}~\bibnamefont{Caruso}},
  \bibinfo{author}{\bibfnamefont{J.}~\bibnamefont{Eisert}},
  \bibinfo{author}{\bibfnamefont{V.}~\bibnamefont{Giovannetti}},
  \bibnamefont{and} \bibinfo{author}{\bibfnamefont{A.~S.}
  \bibnamefont{Holevo}}, \bibinfo{journal}{New Journal of Physics}
  \textbf{\bibinfo{volume}{10}}, \bibinfo{pages}{083030}
  (\bibinfo{year}{2008}),
  \urlprefix\url{http://stacks.iop.org/1367-2630/10/i=8/a=083030}.

\bibitem[{\citenamefont{Holevo and Giovannetti}(2012)}]{Holevo2012}
\bibinfo{author}{\bibfnamefont{A.~S.} \bibnamefont{Holevo}} \bibnamefont{and}
  \bibinfo{author}{\bibfnamefont{V.}~\bibnamefont{Giovannetti}},
  \bibinfo{journal}{Reports on Progress in Physics}
  \textbf{\bibinfo{volume}{75}}, \bibinfo{pages}{046001}
  (\bibinfo{year}{2012}),
  \urlprefix\url{http://stacks.iop.org/0034-4885/75/i=4/a=046001}.

\bibitem[{\citenamefont{Simon}(2000)}]{Simon2000}
\bibinfo{author}{\bibfnamefont{R.}~\bibnamefont{Simon}},
  \bibinfo{journal}{Phys. Rev. Lett.} \textbf{\bibinfo{volume}{84}},
  \bibinfo{pages}{2726} (\bibinfo{year}{2000}),
  \urlprefix\url{http://link.aps.org/doi/10.1103/PhysRevLett.84.2726}.

\bibitem[{\citenamefont{\and Khalili~F.}(1992)}]{Braginsky1992}
\bibinfo{author}{\bibfnamefont{B.~V.} \bibnamefont{\and Khalili~F.}},
  \emph{\bibinfo{title}{Quantum Measurement}} (\bibinfo{publisher}{Cambridge
  University Press}, \bibinfo{year}{1992}).

\bibitem[{\citenamefont{{Heisenberg}}(1983)}]{Heisenberg1927}
\bibinfo{author}{\bibfnamefont{W.}~\bibnamefont{{Heisenberg}}},
  \bibinfo{journal}{Zhurnal Physik} \textbf{\bibinfo{volume}{43}},
  \bibinfo{pages}{172} (\bibinfo{year}{1983}).

\bibitem[{\citenamefont{Arthurs and Kelly~Jr.}(1965)}]{arthurs1965}
\bibinfo{author}{\bibfnamefont{E.}~\bibnamefont{Arthurs}} \bibnamefont{and}
  \bibinfo{author}{\bibfnamefont{J.~L.} \bibnamefont{Kelly~Jr.}},
  \bibinfo{journal}{Bell System Technical Journal}
  \textbf{\bibinfo{volume}{44}}, \bibinfo{pages}{725} (\bibinfo{year}{1965}).

\bibitem[{\citenamefont{Busch et~al.}(2013)\citenamefont{Busch, Lahti, and
  Werner}}]{Busch2013}
\bibinfo{author}{\bibfnamefont{P.}~\bibnamefont{Busch}},
  \bibinfo{author}{\bibfnamefont{P.}~\bibnamefont{Lahti}}, \bibnamefont{and}
  \bibinfo{author}{\bibfnamefont{R.~F.} \bibnamefont{Werner}}
  (\bibinfo{year}{2013}), 
  \eprint{1306.1565}, \urlprefix\url{http://arxiv.org/abs/1306.1565}.

\bibitem[{\citenamefont{Page and Geilker}(1981)}]{page.1981.979--982}
\bibinfo{author}{\bibfnamefont{D.~N.} \bibnamefont{Page}} \bibnamefont{and}
  \bibinfo{author}{\bibfnamefont{C.~D.} \bibnamefont{Geilker}},
  \bibinfo{journal}{Physical Review Letters} \textbf{\bibinfo{volume}{47}},
  \bibinfo{pages}{979} (\bibinfo{year}{1981}),
  \urlprefix\url{http://link.aps.org/doi/10.1103/PhysRevLett.47.979}.

\bibitem[{\citenamefont{Milatz et~al.}(1953)\citenamefont{Milatz, Zolingen, and
  Iperen}}]{Milatz1953}
\bibinfo{author}{\bibfnamefont{J.}~\bibnamefont{Milatz}},
  \bibinfo{author}{\bibfnamefont{J.~V.} \bibnamefont{Zolingen}},
  \bibnamefont{and} \bibinfo{author}{\bibfnamefont{B.~V.}
  \bibnamefont{Iperen}}, \bibinfo{journal}{Physica}
  \textbf{\bibinfo{volume}{19}}, \bibinfo{pages}{195 } (\bibinfo{year}{1953}),
  ISSN \bibinfo{issn}{0031-8914},
  \urlprefix\url{http://www.sciencedirect.com/science/article/pii/S0031891453800212}.

\bibitem[{\citenamefont{Mancini et~al.}(1998)\citenamefont{Mancini, Vitali, and
  Tombesi}}]{Mancini1998}
\bibinfo{author}{\bibfnamefont{S.}~\bibnamefont{Mancini}},
  \bibinfo{author}{\bibfnamefont{D.}~\bibnamefont{Vitali}}, \bibnamefont{and}
  \bibinfo{author}{\bibfnamefont{P.}~\bibnamefont{Tombesi}},
  \bibinfo{journal}{Phys. Rev. Lett.} \textbf{\bibinfo{volume}{80}},
  \bibinfo{pages}{688} (\bibinfo{year}{1998}),
  \urlprefix\url{http://link.aps.org/doi/10.1103/PhysRevLett.80.688}.

\bibitem[{\citenamefont{Cohadon et~al.}(1999)\citenamefont{Cohadon, Heidmann,
  and Pinard}}]{Cohadon1999}
\bibinfo{author}{\bibfnamefont{P.~F.} \bibnamefont{Cohadon}},
  \bibinfo{author}{\bibfnamefont{A.}~\bibnamefont{Heidmann}}, \bibnamefont{and}
  \bibinfo{author}{\bibfnamefont{M.}~\bibnamefont{Pinard}},
  \bibinfo{journal}{Phys. Rev. Lett.} \textbf{\bibinfo{volume}{83}},
  \bibinfo{pages}{3174} (\bibinfo{year}{1999}),
  \urlprefix\url{http://link.aps.org/doi/10.1103/PhysRevLett.83.3174}.

\bibitem[{\citenamefont{Abbott et~al.}(2009)\citenamefont{Abbott, Abbott,
  Adhikari, Ajith, Allen, Allen, Amin, Anderson, Anderson, Arain
  et~al.}}]{LIGO2009}
\bibinfo{author}{\bibfnamefont{B.}~\bibnamefont{Abbott}},
  \bibinfo{author}{\bibfnamefont{R.}~\bibnamefont{Abbott}},
  \bibinfo{author}{\bibfnamefont{R.}~\bibnamefont{Adhikari}},
  \bibinfo{author}{\bibfnamefont{P.}~\bibnamefont{Ajith}},
  \bibinfo{author}{\bibfnamefont{B.}~\bibnamefont{Allen}},
  \bibinfo{author}{\bibfnamefont{G.}~\bibnamefont{Allen}},
  \bibinfo{author}{\bibfnamefont{R.}~\bibnamefont{Amin}},
  \bibinfo{author}{\bibfnamefont{S.~B.} \bibnamefont{Anderson}},
  \bibinfo{author}{\bibfnamefont{W.~G.} \bibnamefont{Anderson}},
  \bibinfo{author}{\bibfnamefont{M.~A.} \bibnamefont{Arain}},
  \bibnamefont{et~al.}, \bibinfo{journal}{New Journal of Physics}
  \textbf{\bibinfo{volume}{11}}, \bibinfo{pages}{073032}
  (\bibinfo{year}{2009}),
  \urlprefix\url{http://stacks.iop.org/1367-2630/11/i=7/a=073032}.

\bibitem[{\citenamefont{Lee et~al.}(2010)\citenamefont{Lee, McRae, Harris,
  Knittel, and Bowen}}]{Lee2010}
\bibinfo{author}{\bibfnamefont{K.~H.} \bibnamefont{Lee}},
  \bibinfo{author}{\bibfnamefont{T.~G.} \bibnamefont{McRae}},
  \bibinfo{author}{\bibfnamefont{G.~I.} \bibnamefont{Harris}},
  \bibinfo{author}{\bibfnamefont{J.}~\bibnamefont{Knittel}}, \bibnamefont{and}
  \bibinfo{author}{\bibfnamefont{W.~P.} \bibnamefont{Bowen}},
  \bibinfo{journal}{Phys. Rev. Lett.} \textbf{\bibinfo{volume}{104}},
  \bibinfo{pages}{123604} (\bibinfo{year}{2010}),
  \urlprefix\url{http://link.aps.org/doi/10.1103/PhysRevLett.104.123604}.

\bibitem[{\citenamefont{Newman et~al.}(1996)\citenamefont{Newman, Bantel,
  Beilby, Krishnan, Siragusa, Boynton, and Goodson}}]{UW}
\bibinfo{author}{\bibfnamefont{R.~D.} \bibnamefont{Newman}},
  \bibinfo{author}{\bibfnamefont{M.}~\bibnamefont{Bantel}},
  \bibinfo{author}{\bibfnamefont{M.}~\bibnamefont{Beilby}},
  \bibinfo{author}{\bibfnamefont{N.}~\bibnamefont{Krishnan}},
  \bibinfo{author}{\bibfnamefont{E.}~\bibnamefont{Siragusa}},
  \bibinfo{author}{\bibfnamefont{P.~E.} \bibnamefont{Boynton}},
  \bibnamefont{and} \bibinfo{author}{\bibfnamefont{A.}~\bibnamefont{Goodson}},
  in \emph{\bibinfo{booktitle}{7th Marcel Grossman Meeting on General
  Relativity}} (\bibinfo{publisher}{World Scientific}, \bibinfo{year}{1996}),
  p. \bibinfo{pages}{1619}.

\bibitem[{\citenamefont{Bantel and Newman}(2000)}]{bantel.2000.233--242}
\bibinfo{author}{\bibfnamefont{M.}~\bibnamefont{Bantel}} \bibnamefont{and}
  \bibinfo{author}{\bibfnamefont{R.}~\bibnamefont{Newman}},
  \bibinfo{journal}{Journal of Alloys and Compounds}
  \textbf{\bibinfo{volume}{310}}, \bibinfo{pages}{233} (\bibinfo{year}{2000}),
  ISSN \bibinfo{issn}{0925-8388},
  \urlprefix\url{http://www.sciencedirect.com/science/article/pii/S0925838800010100}.

\bibitem[{\citenamefont{Calzetta and Hu}(1994)}]{calzetta.1994.6636--6655}
\bibinfo{author}{\bibfnamefont{E.}~\bibnamefont{Calzetta}} \bibnamefont{and}
  \bibinfo{author}{\bibfnamefont{B.~L.} \bibnamefont{Hu}},
  \bibinfo{journal}{Physical Review D} \textbf{\bibinfo{volume}{49}},
  \bibinfo{pages}{6636} (\bibinfo{year}{1994}),
  \urlprefix\url{http://link.aps.org/doi/10.1103/PhysRevD.49.6636}.

\bibitem[{\citenamefont{Heinosaari et~al.}(2010)\citenamefont{Heinosaari,
  Holevo, and Wolf}}]{Heinosaari2010}
\bibinfo{author}{\bibfnamefont{T.}~\bibnamefont{Heinosaari}},
  \bibinfo{author}{\bibfnamefont{A.~S.} \bibnamefont{Holevo}},
  \bibnamefont{and} \bibinfo{author}{\bibfnamefont{M.~M.} \bibnamefont{Wolf}},
  \bibinfo{journal}{Quantum Info. Comput.} \textbf{\bibinfo{volume}{10}},
  \bibinfo{pages}{619} (\bibinfo{year}{2010}), ISSN \bibinfo{issn}{1533-7146},
  \urlprefix\url{http://dl.acm.org/citation.cfm?id=2011373.2011377}.

\bibitem[{\citenamefont{Werner and Wolf}(2001)}]{Werner2001}
\bibinfo{author}{\bibfnamefont{R.~F.} \bibnamefont{Werner}} \bibnamefont{and}
  \bibinfo{author}{\bibfnamefont{M.~M.} \bibnamefont{Wolf}},
  \bibinfo{journal}{Phys. Rev. Lett.} \textbf{\bibinfo{volume}{86}},
  \bibinfo{pages}{3658} (\bibinfo{year}{2001}),
  \urlprefix\url{http://link.aps.org/doi/10.1103/PhysRevLett.86.3658}.

\end{thebibliography}
\end{document}